\documentclass[preprint,prd,showpacs,amsmath,amssymb,amsthm,nofootinbib]{revtex4}
\usepackage[mathscr]{euscript}
\usepackage{bm}% bold math
\usepackage{graphicx}% Include figure files
\usepackage{subfigure}
\usepackage[colorlinks=true,linkcolor=red]{hyperref}
\usepackage{color}
\usepackage{ulem}
\newcommand{\bea}{\begin{eqnarray}}
\newcommand{\eea}{\end{eqnarray}}
\newcommand{\beq}{\begin{equation}}
\newcommand{\eeq}{\end{equation}}

\def\/{\over}

\begin{document}

\title{Accelerating AdS black holes as the  holographic heat engines in a benchmarking scheme }
\author{Jialin Zhang, Yanjun Li,  and Hongwei Yu\footnote{Corresponding author at hwyu@hunnu.edu.cn} }

\affiliation {Department of Physics and Synergetic Innovation Center for Quantum Effects and Applications, Hunan Normal University, Changsha, Hunan 410081, China}

\begin{abstract}

 We investigate the properties of holographic heat engines  with an uncharged accelerating non-rotating AdS black hole as the working substance in a benchmarking scheme. We find that the efficiencies of the black hole heat engines can be influenced by both the size of  the benchmark circular cycle and  the cosmic string tension as a thermodynamic variable. In general, the  efficiency can be increased by enlarging the  cycle, but is
still constrained by a  universal bound $2\pi/(\pi+4)$  as expected.  A cross-comparison of the efficiencies of the accelerating black hole heat engines and Schwarzschild-AdS black hole heat engines suggests that the acceleration also increases the efficiency although the amount of increase is not remarkable.
\end{abstract}
%%04.70.Dy, 04.50.Gh, 05.70.Ce

\maketitle

\section{Introduction}
Black holes are fascinating objects which provide a useful link to
explore the relationship between general relativity, thermodynamics
and quantum theory. More than four  decades after Hawking's
discovery of  black hole radiation~\cite{S.H 1975,S.H 1976}, the
black hole thermodynamics has   been established and developed into
an important sub-discipline in physics. The study of  black hole
thermodynamics has already shed some light on the nature of quantum
gravity in the lack of a consistent quantum theory of it.  Recently,
it is found that the thermodynamical properties of black holes in
Anti-de Sitter (AdS) space are quite different from those  in flat
or de Sitter(dS) space, which are thermodynamically  stable, on
one hand, and on the other hand, deep insight  has been gained into
some phenomena in strongly coupled  quantum field theories by means
of  AdS/CFT
correspondence~\cite{Maldacena:1997re,Gubser:1998bc,Witten:1998qj,Witten:1998qj2,Aharony}.

So far,  many approaches have been introduced to analyze  black hole thermodynamics,  such as, those of positing the laws of gravitation to be connected with the laws of thermodynamics~\cite{Jacobson:1995,Padmanabhan:2010}, treating the black holes as a  holographically dual system in quantum chromodynamics~\cite{Kovtun:2005} and condensed matter physics~\cite{Hartonll:2007,Hartnoll:2008},  and approaching the thermodynamics of black holes geometrically~\cite{Aman:2003,Quevedo:2008,Ruppeiner:2014,Mansoori:2014,Suresh:2014,Zhang:2015,Zhang-2:2015,Hendi:2015,Quevedo:2016, Banerjee:2017}. More recently, by elevating the negative cosmological constant $\Lambda$ as the pressure and defining the thermodynamic volume satisfying a reverse isoperimetric inequality~\cite{Cvetic:2011,Hennigar:2015} as  the conjugate to the pressure
 in the extended  black hole thermodynamics~\cite{Teitelboim:1985,Caldarelli:2000,Kastor:2009,Cvetic:2011,Dolan:2011,Dolan:2011-2}, some  interesting thermodynamic phenomena and rich phase structures quite analogous to the van der Waals fluids are discovered~\cite{Kubiznak:2017}, and this burgeoning subject, which named as black hole chemistry~\cite{Kubiznak:2015,Mann:2016}, has attracted a lot of attention.

In the context of black hole chemistry,  Johnson proposed the concept of holographic heat engines which can extract work with AdS black holes  used as the work materials in the pressure-volume phase space~\cite{Johnson:2014}. The name of ``holographic" originates from the fact that the cycle represents a journey defined on the space of dual field theories in one dimension lower~\cite{Johnson:2014}. After Johnson's pioneering work,  subsequent studies have generalized  this concept to other black holes~\cite{Belhaj:2015,Caceres:2015,setare:2015,Johnson:2016-1,Johnson:2016-2,Zhang:2016,
Sadeghi:2017,Wei:2017,Hendi:2017,Bhamidipati:2017,Xu:2017,Liu:2017,Mo:2017,Hennigar:2017-1}. More recently, in order to better compare the efficiency of the heat engines with different black holes as  working substances, Chakraborty and Johnson introduced a circular cycle of the heat engine in the $P-V$ phase space to benchmark black hole heat engines~\cite{Chakraborty:2016}. Since the circular cycle of a heat engine is a judicious choice for all working substances without  favoring any species of black holes, it  can be considered as a benchmarking cycle in general.

In this paper, we plan to generalize, in the benchmarking scheme,
the study of  holographic heat engines to the case of AdS black
holes with acceleration. The accelerating black holes are known to
be described by the so-called
$C-$metric~\cite{Kinnersley:1970,Plebanski:1976,Dias:2003,Griffiths:2006},
which  has been used to investigate the pair creation of black
holes~\cite{Dowker:1994}, the splitting  of cosmic
strings~\cite{Gregory:1995,Eardley:1995}, and even to construct the
black ring in 5-dimensional gravity~\cite{Emparan:2002}.  Recently,
Appels et al have  derived the thermodynamics of accelerating black
holes~\cite{Appels:2016} and generalized the results to the case of
varying conical deficits for $C-$ metric~\cite{Appels:2017,Anabalon:2018}.

The paper is organized as follows. We will review the thermodynamics
of accelerating black holes with conical defects in $C-$ metric in
next section. In Section III, we will study the benchmarking
holographic heat engines with  accelerated AdS black holes as
working substances with numerical analysis. We will summarize and
conclude in Section IV.

\section{A slowly accelerating AdS black hole  and its thermodynamics }

Let us now give a brief review of an  uncharged slowly accelerating AdS
black hole  and its thermodynamics. The accelerated AdS black holes
can be described by $C-$metric\cite{Appels:2016,Appels:2017}.
However, in order to have a well-defined temperature for these black
holes, it is appropriated to restrict the acceleration to be  slow
(slowly accelerating $C-$metric) so that the acceleration horizon
can be negated by a negative cosmological constant and only the
black hole horizon exists~\cite{Appels:2017}.

Then, a  slowly accelerating AdS black hole can be described by the following $C-$metric  \cite{Griffiths:2006,Appels:2016,Appels:2017}
\begin{align}\label{c-metric}
ds^{2}=\frac{1}{\Omega^{2}}\bigg[f(r)dt^{2}-\frac{dr^{2}}{f(r)}-r^{2}\Big(\frac{d\theta^{2}}{g(\theta)}
+g(\theta)\sin^{2}\theta\frac{d\phi^{2}}{K^{2}}\Big)\bigg]\;,
\end{align}
where
\begin{align}
f(r)=(1-A^{2}r^{2})\big(1-\frac{2m}{r}\big)+\frac{r^{2}}{\ell^{2}}\;,
\end{align}
\begin{align}
g(\theta)=1+2mA \cos\theta\;,
\end{align}
and the conformal factor  $\Omega=1+A r\cos\theta$.
Here, the parameters $m$ and  $A$ are related to the mass and  the magnitude of acceleration of the black hole respectively,  $K$ characterizes
the conical deficit of the spacetime, and $\ell$ represents the AdS radius. For the slowly accelerating case, technically , if $A\ell<3\sqrt{3}/4\sqrt{2}$, there is only the black hole event horizon  $r_+$  which satisfies $f(r_+)=0$~\cite{Appels:2017}.  It is easy to see that there is a conical deficit in this spacetime which is  unequal at two different poles, and it is this difference of the deficits that produces an overall force that drives the acceleration. In fact, if we require that the angular part of the metric be regular at a pole, then we have
\begin{equation}
K_{\pm}=1\pm2mA\;,
\end{equation}
which indicates clearly that we can not have regularity at both poles if $2mA\neq 0$, and this kind of irregularity along an axis is  precisely a definition of a conical singularity that signals the existence of a cosmic string. Actually, the tension of the string $\mu$ is related to the conical deficit angle  $\delta$ by $\mu=\delta/8\pi$, where
\begin{equation}
\delta=2\pi\big[1-\frac{g(\theta)}{K}\big]\;.
\end{equation}
So, the string tension varies as we move along the axis,  and on the north pole ($\theta_+=0$ ) and the south pole ($\theta_-=\pi$ ),  the cosmic string tensions are given by
\begin{equation} \label{mu}
\mu_\pm=\frac{1}{4}-\frac{g(\theta_\pm)}{4K}=\frac{1}{4}-\frac{1\pm2m A}{4K}\;.
\end{equation}

 In order to avoid the occurrence of negative tension defects, the requirement of $\mu_+\geq0$ is compulsory.  Clearly, it is easy to find out that $\mu_+\leq\mu_{-}\leq1/4$.

%%%%%%%%%%%%
In order to obtain the correct thermodynamics, it has been argued that the normalization of the timelike Killing vector should be appropriately chosen~\cite{Gibbons:2005}. In fact,  such a normalization of the time coordinate can be obtained such that it corresponds to the ``time" of an asymptotic observer,  $\tau=\alpha{t}$,  with $\alpha=\sqrt{1-A^2\ell^2}$\cite{Anabalon:2018}. The black hole mass associated with the normalized time can be found by using the method of conformal completion\cite{Ashtekar:1999,Das:2000},
\begin{equation}M=\alpha\frac{m}{K}\;.\end{equation}
It should be pointed out that $M$ is usually identified with enthalpy rather than internal energy in  extended black hole thermodynamics.
 As usual, the temperature $T$  is given by using the conventional Euclidean method associated with the normalized time $\tau$,
\begin{equation}\label{T-H}
T=\frac{f'(r_+)}{4\alpha\pi}=\frac{m}{2\pi \alpha r_{+}^{2}}+\frac{A^{2}m}{2\alpha\pi}-\frac{A^{2}r_{+}}{2\alpha\pi}+\frac{r_{+}}{2\pi\alpha \ell^{2}}\;.
\end{equation}
And the  entropy $S$ of the accelerating black hole still obeys the area theorem
\begin{equation}\label{S-H}
S=\frac{\pi r_{+}^{2}}{K(1-A^{2}r_{+}^{2})}\;.
\end{equation}
The thermodynamic pressure associated with the cosmological constant in extended black hole thermodynamics reads
\begin{equation} \label{PV}
P=-\frac{\Lambda}{8\pi}=\frac{3}{8\pi \ell^{2}}\;,\quad
\end{equation}
and the thermodynamic volume is
\begin{equation}\label{PV2}
V=\frac{4\pi }{3K{\alpha}}\Big[\frac{ r_{+}^{3}}{(1-A^{2}r_{+}^{2})^{2}}+m A^2\ell^4\Big]\;.
\end{equation}
Then, if we allow the tension of the string to vary, the first law can be straightforwardly derived~\cite{Anabalon:2018}
\begin{equation}\label{firstlaw1}
\delta M=T\delta {S}+V{\delta}P-\lambda_+{\delta}\mu_+-\lambda_-{\delta}\mu_-\;,
\end{equation}
where
\begin{align}
\lambda_\pm=\frac{1}{\alpha}\Big[\frac{r_{+}}{1-A^2r_+^2}-m\Big(1\pm\frac{2A\ell^2}{r_+}\Big)\Big]\;.
\end{align}
Here, $\lambda_\pm$  is defined as a thermodynamic length\cite{Appels:2017,Anabalon:2018}, which is conjugate to the tension $\mu_\pm$.

%%%%%%%%%%%%%%%%%%%%%%%%%%
 As we have seen,   all the thermodynamic variables are the certain combinations  of the solution parameters. A change of solution parameter $A$ may lead to  changes of the thermodynamic mass $M$ and  the cosmic string tensions, so does the change of other solution parameters. Due to the complicated combinations of the thermodynamic variables, it is quite a challenge to obtain the analytic expression for the efficiency of benchmarking black hole heat engines.

\section{Benchmarking Black hole heat engines and efficiency}
For a valid cycle of a holographic heat engine, the efficiency is defined by
\begin{align}
\eta=1-\frac{Q_C}{Q_H}\;,
\end{align}
where $Q_C$ denotes a  net output heat flow in one cycle,  and $Q_H$ represents a net input heat flow.
In general, we can compute the efficiency by keeping the  corresponding  tensions fixed through the heat engine cycle, then the enthalpy $M$ can be considered as a function of $S$ and $P$.

In  the benchmarking scheme of black hole heat engines~\cite{Johnson:2016-1,Hennigar:2017-1}, a circular or elliptical cycle has been suggested in order to allow for  cross-comparison of the efficiencies of holographic heat engines with different black holes as  working substances.
The  cycle is described by  the following parameterized equation  in the $P-V$ plane,
\begin{align}\label{PVequation}
P(\theta)=P_{0}(1+p\sin\theta)\,, ~~~V(\theta)=V_{0}(1+v\cos\theta)\;,
\end{align}
where $(V_0,P_0)$ is the  center of the cycle. Strictly speaking,  the precise shape of the closed contour in  the $P-V$ plane should be elliptical since the units of $P$  and $V$  are different from each other.

For the special case of $C_V=0$,  i.e., the specific heat capacity at constant volume is vanishing, then enthalpy  $M$ can be considered as a function of $V$ and $P$.  By tilling the circular/elliptical cycle with  a series of rectangles,  a net output heat flow can be written in a simple form  by applying Eq.~(\ref{PVequation})
 \begin{align}\label{qccv0}
 Q_C=M(V_0(1+v),P_0)-M(V_0(1-v),P_0)-\frac{\pi P_0V_0pv}{2}\;.
 \end{align}
Performing a similar algorithmic manipulation, we can also get $Q_H$. Then the efficiency of the benchmarking black hole heat engine can be calculated  exactly. However, for rotating  black holes or the case with multiple complicated thermodynamic variables, it is not easy to exactly determine which part of the cycle curve is related to the  output heat flow, then Eq.~(\ref{qccv0}) may be inapplicable to determining the net output heat flow,  and a numerical integration of $TdS$ will be needed~\cite{Hennigar:2017-1}.

For convenience, we can adjust the unit of each coordinate axis in the $P-V$ plane so that the  shape of the cycle contour looks like a circle with a radius of a numerical value $R$. Then the circular  cycle Eq.~(\ref{PVequation}) reads
\begin{align}\label{PV-2}
P(\theta)=P_{0}+R\sin\theta\;, V(\theta)=V_{0}+R\cos\theta\;.
\end{align}
In order to keep the pressure and volume positive on the circular cycle, the radius $R$ needs to be constrained by the circle center $P_0$ and $V_0$, i.e., $R<P_0$ and $R<V_0$.
In the following, we will examine properties of heat engines of an accelerated AdS black hole by numerical estimations.

\begin{figure}[htbp]
\centering{\subfigure[]{\label{circle11}
\includegraphics[width=0.4\textwidth,height=0.4\textwidth]{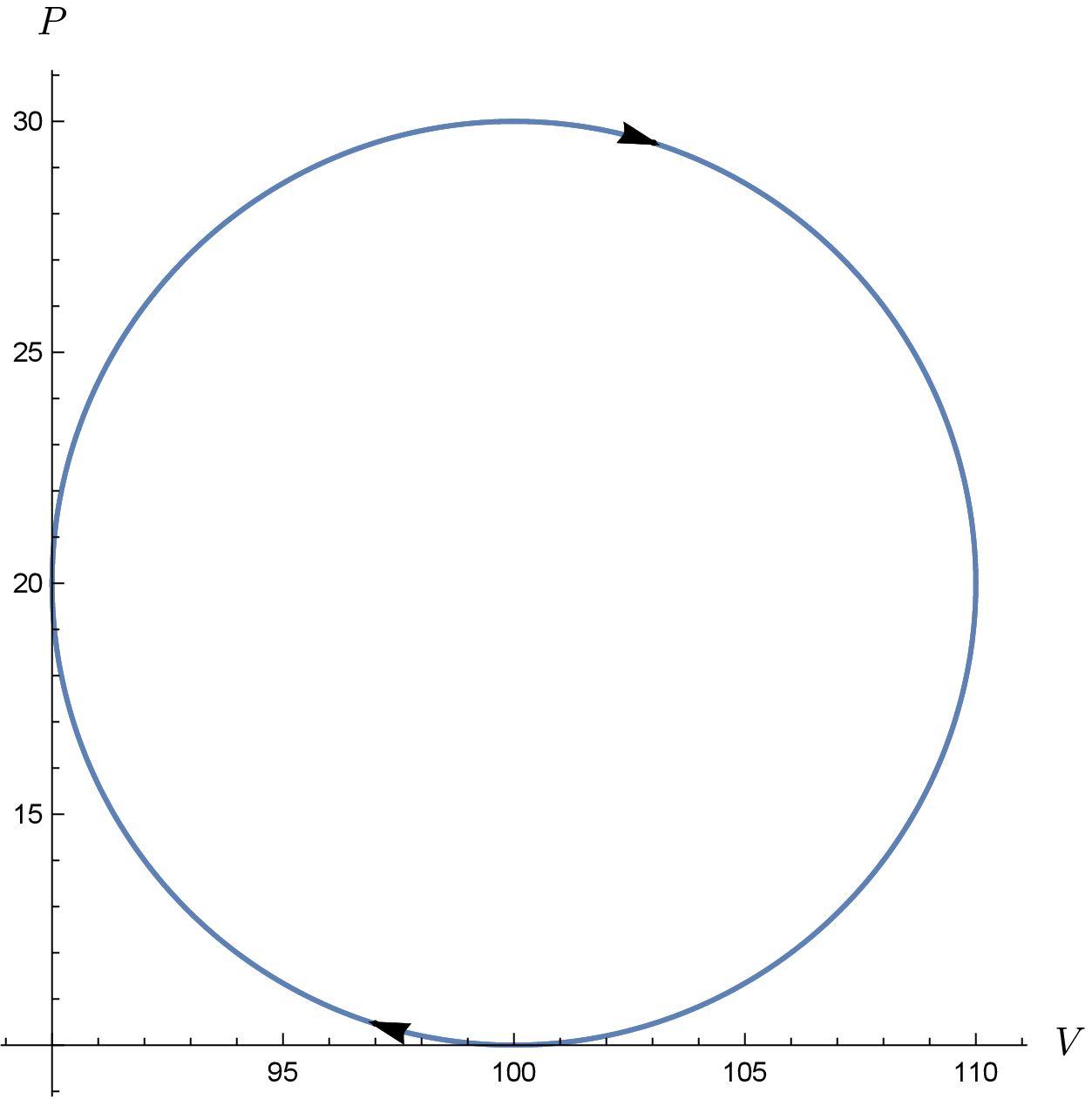}}\subfigure[]{\label{circle22}
\includegraphics[width=0.4\textwidth,height=0.4\textwidth]{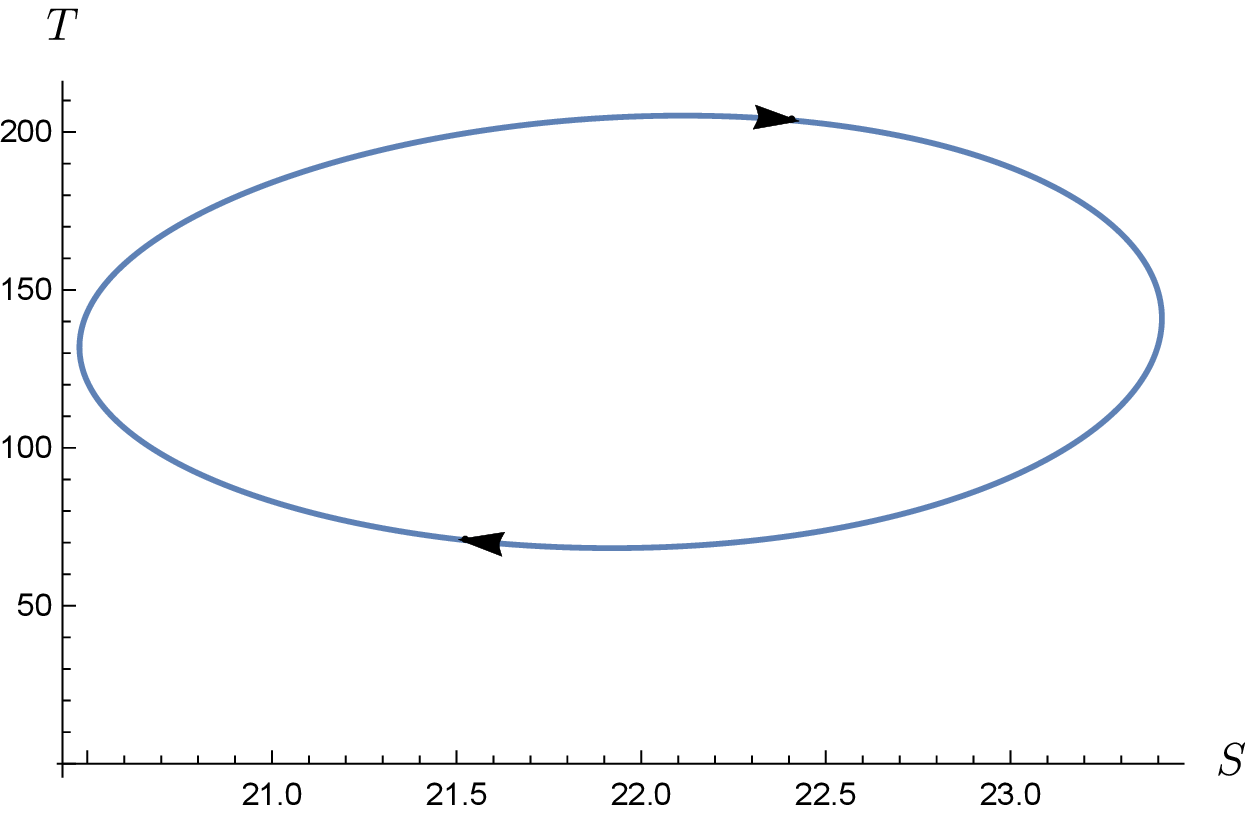}}}
\caption{ For fixed tension $\mu_-=0.2$, the corresponding cycle diagram of an accelerating black hole heat engine  is respectively plotted in (a) $P-V$ plane and (b) $T-S$ plane.  Here, the center of the circle in $P-V$ plane is $(V_0,P_0)=(100,20)$ with the radius $R=10$.   }\label{circle}
\end{figure}

Now, without loss of generality,  we  choose
$K=K_+=g(\theta_+)=1+2m A$ in Eq.~(\ref{mu}) so that the metric is regular on
the north pole (i.e., $\mu_+=0$), then the cosmic string tension on
the south pole becomes
\begin{align}
\mu_-=\frac{K-1}{2K}\;.
\end{align}
%%%%%%%%%%%%%%
For fixed tension,  the corresponding  cycle contour in the $T-S$ plane  can also be directly obtained  by some algebraic manipulations  of the  thermodynamic variables (see Fig.~(\ref{circle})).
According to Eq.~(\ref{S-H}) and Eq.~(\ref{PV2}), it is not difficult to deduce that $C_V\neq0$ in the case of accelerated black holes. Therefore, the numerical treatment should be called for.

\begin{figure}[htbp]
\centering{
\includegraphics[width=0.7\textwidth]{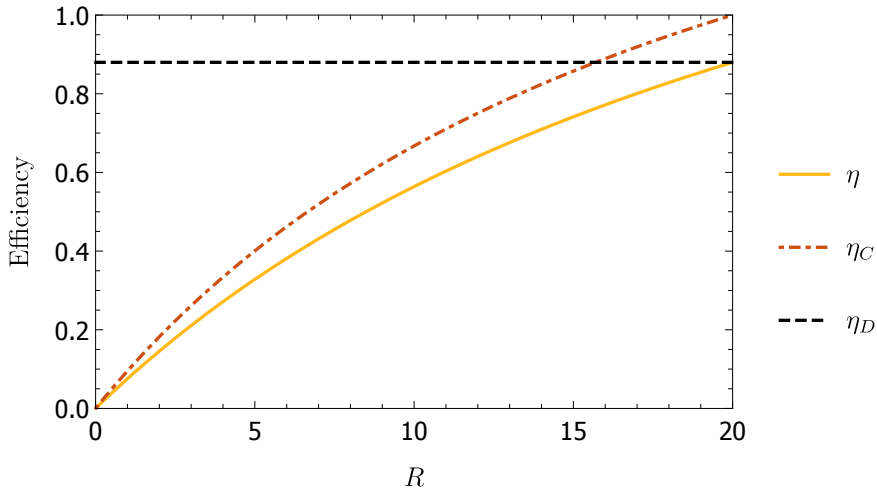}}
\caption{ The efficiency of  a heat engine of  accelerating black holes is plotted as a function of radius $R$.  Here,  we assume the center of the circular cycle in $P-V$ plane is $(V_0,P_0)=(100,20)$ with the cosmic string tension  $\mu_-=0.2$, then the radius for a valid heat engine cycle should satisfy $R<20$.  Note that $\eta_C$ denotes Carnot efficiency and $\eta_{D}$ means the upper-bound value of efficiency.  }\label{carnot-eta}
\end{figure}

In Fig.~(\ref{carnot-eta}), we display the relation between the efficiencies of holographic heat engines and the size of the circular cycle in the $P-V$ plane. Here, the Carnot efficiency  is defined by $\eta_C=1-T_C/T_H$ with maximum temperature $T_H$ and minim temperature $T_C$ in the entire cycle process.
 We can see  from the figure that  both the efficiency $\eta$ and Carnot efficiency $\eta_C$ grow with the increasing radius $R$  and
 the Carnot efficiency is always higher than the efficiency of benchmarking black hole heat engines. Furthermore, the larger the radius of the circle, the bigger the difference. Therefore,  the efficiency $\eta$  cannot approach the Carnot efficiency $\eta_C$ even when the working area of the $P-V$ plane becomes larger. Besides these, it is worth noting that the universal upper-bound  proposed in Ref.~\cite{Hennigar:2017-1} is still valid, i.e.,
\begin{equation}
\eta\leq\eta_D=\frac{2\pi}{\pi+4}\;,
\end{equation}
and the efficiency $\eta$  approaches the upper-bound in the limit of $R\rightarrow {P_0}$ .

\begin{figure}[hbp]
\centering{
\includegraphics[width=0.7\textwidth]{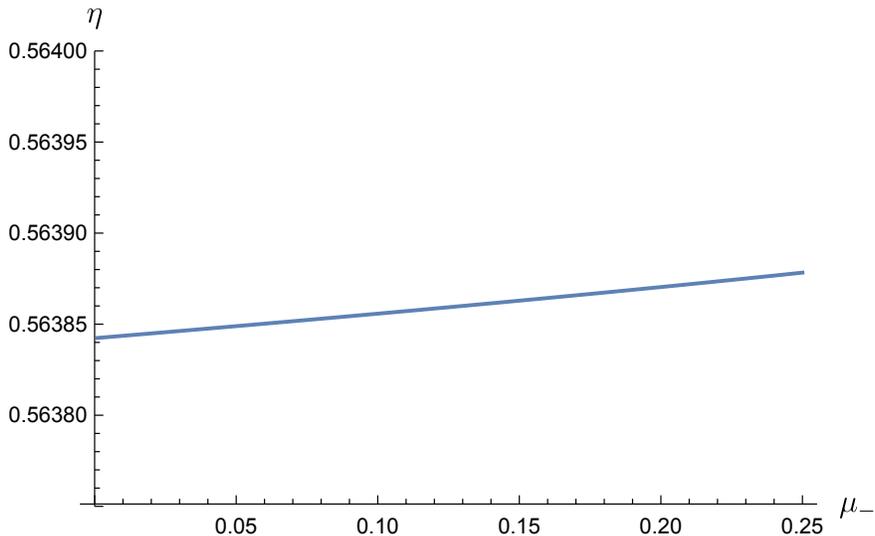}}
\caption{ The efficiency of a heat engine of  accelerating black holes  is plotted as a function of  $\mu_-$.  Here, the circular cycle  is assumed  to satisfy $R=10\;$ with origin at $(V_0,P_0)=(100,20)$. }\label{q0etamu}
\end{figure}

For a fixed circular cycle radius, we have plotted the  benchmarking heat engine efficiency  as a function of the cosmic string tension in Fig.~(\ref{q0etamu}),  which shows that  a stronger  cosmic string tension will  in principle lead to  a somewhat higher efficiency, but  the change in efficiency is actually insignificant since the difference shows up only in the 5th significant figure.

\begin{figure}[ht]
\centering{\subfigure[]{\label{effi-R11}
\includegraphics[width=0.51\textwidth]{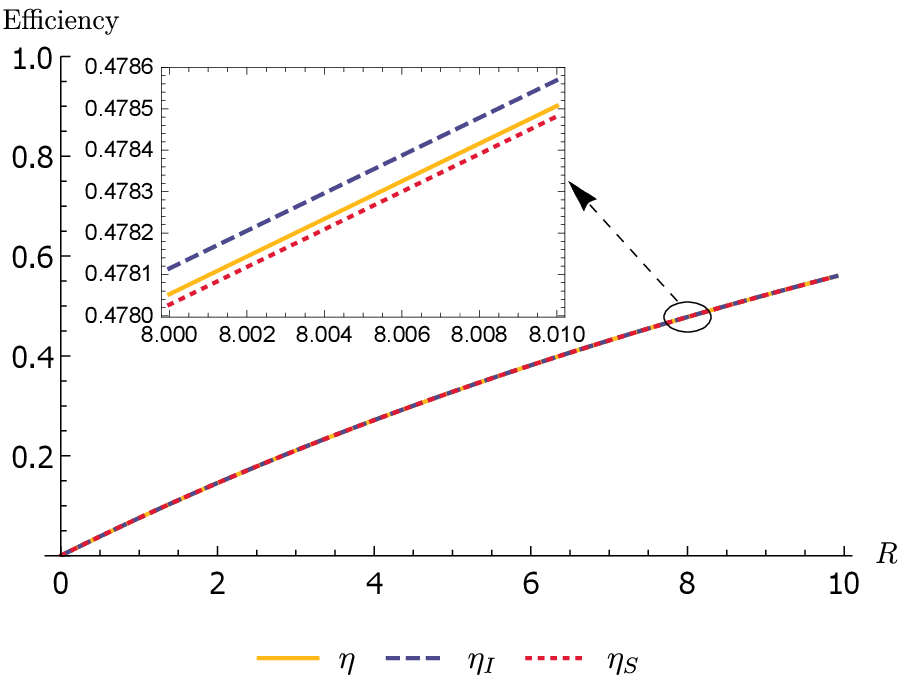}}\subfigure[]{\label{effi-R22}
\includegraphics[width=0.51\textwidth]{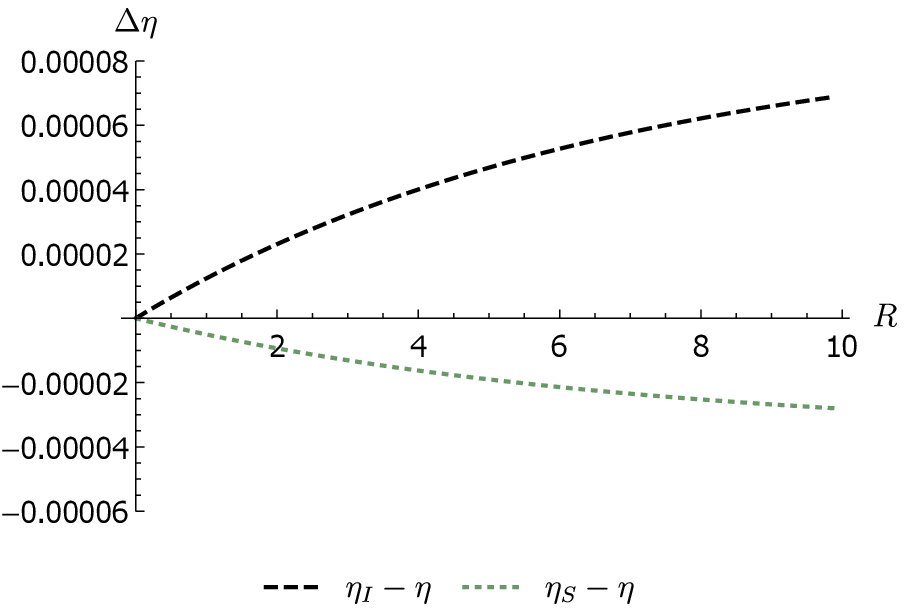}}}
\caption{(a)The efficiencies vs the radius $R$ for three different families of black holes as the working material, including accelerating AdS (denoted by $\eta$ with $\mu_-=0.2$), the ideal gas (denoted by $\eta_I$) and Schwarzschild-AdS (denoted by $\eta_S$) black holes. The differences in the efficiencies are described in (b), where $\eta_I-\eta$ and $\eta_S-\eta$ are represented by the black dashed line  and the green dotted line respectively.  Note that  all the benchmarking circular cycle is centered at $(V_0,P_0)=(100,20)$.}\label{effi-R-all}
\end{figure}

%%%%%%%%%%%%%
 Now, we cross-compare  the efficiency of different black hole heat engines in the benchmarking scheme.  The black hole  heat engines we choose are those of the slowly accelerating AdS black holes we just studied, the ideal gas black holes and the Schwarzschild-AdS black holes. We plot the efficiency of the benchmarking heat  engine as a function of the radius $R$  with different black holes as working substances  in  Fig.~(\ref{effi-R-all}). It is worth noting that both the ideal gas black hole  and Schwarzschild-AdS black hole have a vanishing specific heat at constant volume (i.e., $C_V=0$, the detailed thermodynamic quantities can be found in Ref.~\cite{Kubiznak:2017,Chakraborty:2016,Hennigar:2017-1}), and the  holographic heat engine efficiency, which can be obtained directly by using Eq.~(\ref{qccv0}) in the benchmarking scheme,  has been studied in  Ref.~\cite{Hennigar:2017-1}.

Fig.~(\ref{effi-R-all}) reveals that the work efficiency of holographic heat engines of slowly accelerating black holes, in general, is larger than that of  Schwarzschild-AdS black holes, but smaller than that of ideal gas black holes. Since in the zero acceleration limit the metric (\ref{c-metric}) approaches that of a  Schwarzschild-AdS black hole, one may conclude that with acceleration, the holographic heat engine efficiency is usually larger than that without although the difference is tiny.

\section{Conclusion}
We have explored the properties of holographic heat engines with accelerating  black holes as the working substances. This family of accelerating  AdS black holes described by $C-$ metric represents a black hole with conical deficits along one axis.  These conical deficits provide a driving force to generate the acceleration of a black hole. Physically, the topological defect originated from the conical deficit can be interpreted as a finite-width cosmic string core. Due to the fact that all thermodynamic variables of the accelerating black holes are non-linear combinations of the solution parameters $r_+$, $ m$, $\ell$ and $A$,  it is quite a challenge to obtain an analytical expression for the efficiency of the holographic heat engines and numerical estimations are resorted to.

 In a benchmarking scheme, we have examined the influence of the size of a benchmarking circular cycle and the cosmic string tension on the efficiency of black hole heat engines.  We find that the efficiency can be increased by  enlarging the cycle, but the efficiency cannot exceed the Carnot efficiency as  we would expect. Moreover, it is also constrained by a universal bound  $2\pi/(\pi+4)$ proposed in  Ref.~\cite{Hennigar:2017-1}.  When  the cosmic string tension is varied, the efficiency in principle increases  with the increasing cosmic string tension, but the amount of increase is tiny.

A cross-comparison of the holographic heat engines with slowly accelerating  AdS black holes and the Schwarzschild-AdS black holes in a same benchmarking cycle shows that the presence of acceleration also increases the efficiency somewhat.
Finally, it is worth pointing out that  the thermodynamics and the behavior of holographic heat
engines, when the  accelerating black hole is charged, is an interesting but non-trivial issue, which we would rather leave to a future work.

\begin{acknowledgments}
 This work was supported by the National Natural Science Foundation of China under Grants  No. 11435006 and No.11690034.
\end{acknowledgments}

\end{document}